\documentclass[12pt,fleqn]{article}

\usepackage{graphicx}
\usepackage{pdftricks}
\usepackage{lineno}
\usepackage{dcolumn}

\begin{document}

\title{\Large Unitary thermodynamics from thermodynamic geometry II: Fit to a local density approximation}

\author{
   George Ruppeiner\footnote{ruppeiner@ncf.edu}\\
   Division of Natural Sciences\\
   New College of Florida\\
   5800 Bay Shore Road\\
   Sarasota, Florida 34243-2109}

\maketitle

\begin{abstract} Strongly interacting Fermi gasses at low density possess universal thermodynamic properties that have recently seen very precise $PVT$ measurements by a group at MIT.  This group determined local thermodynamic properties of a system of ultra cold $^6\mbox{Li}$ atoms tuned to Feshbach resonance.  In this paper, I analyze the MIT data with a thermodynamic theory of unitary thermodynamics based on ideas from critical phenomena.  This theory was introduced in the first paper of this sequence, and characterizes the scaled thermodynamics by the entropy per particle $z= S/N k_B$, and the energy per particle $Y(z)$, in units of the Fermi energy.  $Y(z)$ is in two segments, separated by a second-order phase transition at $z=z_c$: a ``superfluid'' segment for $z<z_c$, and a ``normal'' segment for $z>z_c$.  For small $z$, the theory obeys a series $Y(z)=y_0+y_1 z^{\alpha }+y_2 z^{2 \alpha}+\cdots,$ where $\alpha$ is a constant exponent, and $y_i$ ($i\ge 0$) are constant series coefficients.  For large $z$, the theory obeys a perturbation of the ideal gas $Y(z)= \tilde{y}_0\,\mbox{exp}[2\gamma z/3]+ \tilde{y}_1\,\mbox{exp}[(2\gamma/3-1)z]+ \tilde{y}_2\,\mbox{exp}[(2\gamma/3-2)z]+\cdots$ where $\gamma$ is a constant exponent, and $\tilde{y}_i$ ($i\ge 0$) are constant series coefficients.  This limiting form for large $z$ differs from the series used in the first paper, and was necessary to fit the MIT data.  I fit the MIT data by adjusting four free independent theory parameters: $(\alpha,\gamma,\tilde{y}_0,\tilde{y}_1)$.  This fit process was augmented by trap integration and comparison with earlier thermal data taken at Duke University.  The overall match to both the data sets was good, and had $\alpha=1.21(3)$, $\gamma=1.21(3)$, $z_c=0.69(2)$, scaled critical temperature $T_c/T_F=0.161(3)$, where $T_F$ is the Fermi temperature, and Bertsch parameter $\xi_B=0.368(5)$.  I also discuss the virial expansion in the context of this thermodynamic geometric theory.
\end{abstract}


\section{Introduction}

Systems of degenerate, strongly interacting fermions at low density have been of much recent interest as possible models for quark-gluon plasmas, neutron star matter, and high temperature superconductors \cite{Luo2007, Horikoshi2010}.  Model systems of this type are readily prepared in low temperature optical traps, with magnetic fields tuned to produce states near Feshbach resonance \cite{Chin2010}.  The expectation is universal thermodynamic properties, identical for all systems belonging to such a class of systems \cite{Ho2004a}.

\par
In a recent paper \cite{Ruppeiner2014a}, I proposed that this unitary thermodynamics is a good candidate for evaluation by thermodynamic metric geometry, related to ideas from critical phenomena.  The result is a full thermodynamic fundamental equation for all of the thermodynamic properties.  This Local Density Approximation (LDA) contains a few undetermined parameters that may be determined from fits to available experimental data.  Previously \cite{Ruppeiner2014a}, I analyzed thermodynamic data for the total entropy and the total energy taken at Duke University in a nonhomogeneous trapped system \cite{Luo2007, Luo2009}.  A complication in analyzing this data was that trap integration of the theoretical LDA was required to compare with experiment.

\par
In this paper, I fit a somewhat modified version of this thermodynamic geometric LDA to a more recent experiment, at MIT \cite{Ku2012}, which measured local thermodynamic properties, and that required no trap integration of the theory for comparison.  In addition, I trap integrated my resulting fit to the MIT data to compare with the earlier Duke experiment \cite{Luo2007,Luo2009}, and this comparison has led to a recommendation for the best overall fit.  This best overall fit agrees with the MIT data sets with a value of $\chi^2$ of about $1.7$, and with the Duke data with $\chi^2$ of about $0.5$.

\par
I also include three Appendices: Appendix A presents the series solution for $Y_H(z)$, Appendix B presents a virial expansion solution, and Appendix C gives an explicit formula for the best overall fit, and for a number of thermodynamic quantities.

\section{Thermodynamic geometric theory}

In this section, I summarize the general theory, omitting some details found in the first paper \cite{Ruppeiner2014a}.

\subsection{General considerations}

\par
The application of statistical mechanics to unitary thermodynamics runs into difficulties on two points.  First, even if the interaction potential between atoms were exactly known, the calculation of the partition function $Z$ would be very difficult.  Strong interactions typically produce substantial organized fluctuating structures containing many particles.  These mesoscopic structures, of size the correlation length $\xi$, do not lend themselves to familiar approximation schemes such as mean field theory.  Second, even if we could somehow overcome these obstacles for some given system, establishing universality is awkward, since the calculation would have to be repeated for a set of systems, all presumably in the same universality class, to demonstrate equivalence.

\par
A thermodynamic approach offers another way to address such problems.  Thermodynamics requires no specific input of an interparticle interaction potential.  In addition, large organized fluctuating mesoscopic structures are conceptually an advantage, since thermodynamic averages improve with the addition of particles.  However, to exploit such possible thermodynamic advantages requires special thermodynamic tools for dealing with mesoscopic fluctuating structures.  The first such tool is the thermodynamic curvature $R$, that in cases where $\xi$ encompasses a number of particles, is the correlation volume: $|R|\sim\xi^3$.  (For reviews, see \cite{Ruppeiner1995,Ruppeiner2014,Johnston2003}).  The second tool is hyperscaling from the theory of critical phenomena \cite{Widom1974}, which has free energy per volume $\phi\sim\xi^{-3}$.  Eliminating the common $\xi^3$ yields the geometric equation

\begin{equation}R=-\frac{\kappa}{\phi},\label{10}\end{equation}

\noindent where $\kappa$ is a dimensionless constant of order unity.

\par
This geometric equation constitutes a third-order partial differential equation for $\phi$.  In applications so far, however, a scaling assumption of some form was used to express the geometric equation as an ordinary differential equation for a single function.  This ordinary differential equation may be solved, subject to background subtraction, boundary conditions, and assumptions about analyticity.

\par
\begin{figure}[tpg]
\centering
\includegraphics[width=4.0in]{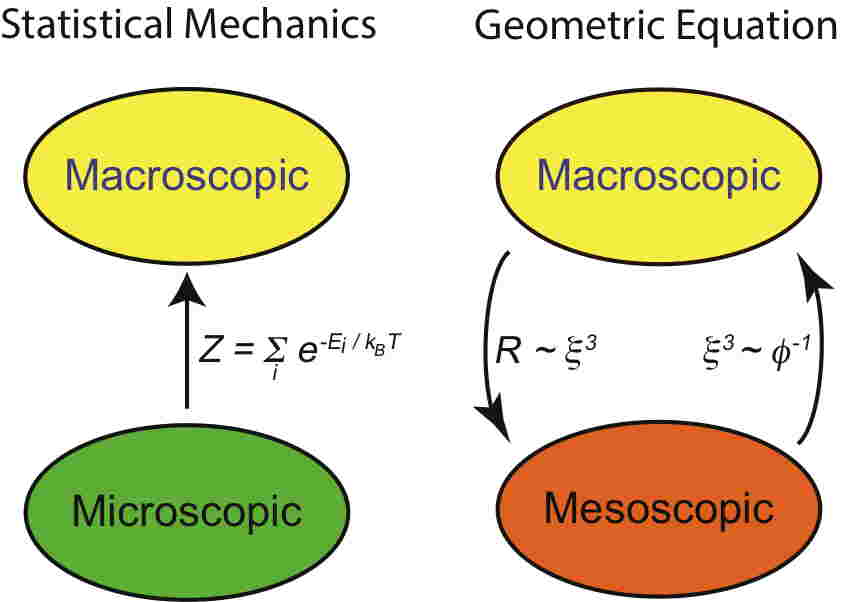}
\caption{Statistical mechanics builds up from the microscopic to the macroscopic by calculating the partition function $Z$ from the quantum energy levels $E_i$ and the temperature $T$.  But this direct calculation method is difficult to implement in the presence of strong interparticle interactions.  Instead, I suggest a thermodynamic calculation method based on the interplay with mesoscopic structures of characteristic size $\xi$, and employing the thermodynamic curvature $R$ and the free energy per volume $\phi$.  This calculation method highlights universal properties.}
\label{fig:1}
\end{figure}

The basic calculation scheme is shown in Figure \ref{fig:1}.  In contrast to statistical mechanics, which builds up from the microscopic level to the macroscopic level by calculating the partition function $Z$, the thermodynamic approach is based on the interplay between the macroscopic and the mesoscopic levels.  The thermodynamic method thus lacks full microscopic information, and certainly cannot hope to encompass all the problems accessible to statistical mechanics.  However, in cases with strong interactions, characterized by universal behavior, the thermodynamic method may offer significant advantages.

\subsection{The scaled form of the internal energy}

Callen \cite{Callen1985} showed that the thermodynamic formalism may be applied in several forms, all corresponding to the same physical results.  In the previous paper \cite{Ruppeiner2014a} I started from the fundamental equation for the internal energy:

\begin{equation} E=E(S, N, V), \label{20}\end{equation}

\noindent where $S$, $N$, and $V$ are the entropy, particle number, and volume, respectively.  Also, let $T=E_{,S}$, $\mu=E_{,N}$, and $p=-E_{,V}$ denote the temperature, chemical potential, and pressure, respectively.  Here, the comma notation indicates differentiation.  In unitary thermodynamics, we use the scaled form:

\begin{equation} E=N \left(\frac{N}{V}\right)^a Y\left[\left(\frac{S}{V}\right)\left(\frac{N}{V}\right)^{-b}\right], \label{30} \end{equation}

\noindent where $a$ and $b$ are constants, and $Y()$ is a function of a single variable.  For the three-dimensional problem, the literature sets $\{a,b\}=\{2/3,1\}$, values corresponding to the ideal Fermi gas \cite{Pathria1996}.  In this case Eq. (\ref{30}) becomes

\begin{equation} E=N\epsilon_F(\rho) Y(z), \label{60} \end{equation}

\noindent where {\cite{Pathria1996}}

\begin{equation} \epsilon_F(\rho) =\left(\frac{3^{2/3}\pi^{4/3}\,\hbar^2}{2m}\right)\rho^{2/3}\label{70}\end{equation}

\noindent is the Fermi energy, $\rho=N/V$ is the particle density,

\begin{equation}z=\frac{S}{N k_B},\label{80}\end{equation}

\noindent is the entropy per particle, $\hbar$ is Planck's constant $h$ divided by $2\pi$, and $m$ is the particle mass.

\par
This scaled form for the energy considerably simplifies the solution of the geometric equation Eq. (\ref{10}).

\subsection{Solutions about $z=0$ and $z\to\infty$}

The thermodynamic geometric theory is based around two singular points, $z\to 0$, where $R\to\infty$, and $z\to\infty$, where $R\to 0$ \cite{Ruppeiner2014a}.  I solved for $Y(z)$ about both singular points, and joined the solutions at $z=z_c$.  This joining cannot be continuous in all the thermodynamic quantities, and I joined to get a second-order phase transition at $z_c$.  A joining corresponding to a first-order phase transition was also possible, but since the MIT group \cite{Ku2012} featured a second-order phase transition in their analysis, I did likewise.  The result was two functional branches $Y_S(z)$ for $z<z_c$, and $Y_H(z)$ for $z>z_c$.

\par
The small $z$, or ``superfluid'', function $Y_S(z)$ was found in the first paper \cite{Ruppeiner2014a}, and for $z$ near zero takes the form of a Puisseux series:

\begin{equation} Y_S(z)=y_0 + y_1 \, z^{\alpha} + y_2 \, z^{2 \alpha} + \cdots. \label{90}\end{equation}

\noindent Here, $\alpha$ is a constant  exponent, and $y_0$, $y_1$, $\cdots$, are series coefficients.  Thermodynamic stability for $z\to 0$ requires $\alpha>1$, $y_0>0$, and $y_1>0$.  $Y_S(z)$ satisfies a third-order differential equation, with three free constants $(\alpha,y_0,y_1)$ determined by data fitting, and with the remaining series coefficients $y_i$ ($i\ge 2$) determined uniquely by series solution of the differential equation for $Y_S(z)$.

\par
In the first paper \cite{Ruppeiner2014a}, I tried a Puisseux series solution as well for $Y_H(z)$ for large $z$, and this proved effective for trap integrating to fit the Duke experiment \cite{Luo2007,Luo2009}.  However, the MIT experiment \cite{Ku2012} has smaller error bars and considerably more data for higher $z$, and a Puisseux series for $Y_H(z)$ will not produce an acceptable fit.  A fundamentally new solution for $Y_H(z)$ is required.

\par
In Appendix A, I demonstrate that a series of the form of a perturbation around the ideal gas solves the geometric equation for large $z$:

\begin{equation} Y_H(z)= \tilde{y}_0\,\mbox{exp}[2\gamma z/3]+ \tilde{y}_1\,\mbox{exp}[(2\gamma/3-1)z]+ \tilde{y}_2\,\mbox{exp}[(2\gamma/3-2)z]+\cdots, \label{100}\end{equation}

\noindent where $\gamma\,(>0)$ is a free constant, determined by data fitting.  The series coefficients $\tilde{y}_0\,(>0)$ and $\tilde{y}_1$ are also free constants.  The remaining series coefficients $\tilde{y}_i$ ($i\ge 2$) are determined in terms of $(\gamma,\tilde{y}_0,\tilde{y}_1)$ from the series solution of the differential equation for $Y_H(z)$.  $\gamma=1$ corresponds to a limiting ideal gas solution (the Sackur-Tetrode equation \cite{Callen1985}, with an appropriate multiplier) as $z$ gets large.

\subsection{Joining of the curves at $z=z_c$}

For a phase transition of any order, we require continuous $\{T,\mu, p\}$.  A second-order phase transition requires, in addition, continuous $\rho$ and $z$.  As was shown in the first paper \cite{Ruppeiner2014a}, these continuity conditions lead to:

\begin{equation} Y_S(z_c)=Y_H(z_c), \label{110} \end{equation}

\noindent and

\begin{equation} Y'_S(z_c) = Y'_H(z_c). \label{120} \end{equation}

\noindent In my approach, neither $Y_S(z)$ nor $Y_H(z)$ show any singular behavior at $z_c$.  Hence, this second-order phase transition corresponds to a pure Ehrenfest phase transition, with discontinuities but no infinities in the second derivative of $Y(z)$.  It is generally thought that the phase transition in unitary thermodynamics is second-order \cite{Parish2007}, and in the 3D XY universality class \cite{Haussmann2008}.  Systems in the 3D XY universality class are not thought to have infinities in their response functions, consistent with the picture developed in this paper.  The MIT experiment \cite{Ku2012} also does not show apparent singular behavior in the response functions, though singular behavior could conceivably be rounded off by the finite resolution of the imaging system. 

\section{Fitting theory to experiment}

In this section, I discuss the fitting of the thermodynamic geometric theory to experiment.  The idea of fitting data in unitary thermodynamics to functions with undetermined parameters was featured by Luo and Thomas \cite{Luo2009}.

\subsection{The MIT experiment}

\par
The MIT group \cite{Ku2012} directly measured the local number density $\rho(\vec{r})$ at position $\vec{r}$, where the trap potential is $U(\vec{r})$.  Resulting density profiles enabled these researchers to determine (theory free) a curve for the reduced compressibility $\tilde{\kappa}=\kappa/\kappa_0$ in terms of the reduced pressure $\tilde{p}=p/p_0$.  Here, the compressibility is

\begin{equation} \kappa = -\frac{1}{V} \left(\frac{\partial V}{\partial p} \right)_T, \label{130} \end{equation}

\noindent and $(\kappa_0, p_0)$ denote the compressibility and the pressure, respectively, of the corresponding zero-temperature noninteracting Fermi gas.  Also define the reduced temperature $\tilde{T}=T/T_F$, where the Fermi temperature $T_F=\epsilon_F(\rho)/k_B$.

\par
The scaled fundamental equation Eq. (\ref{60}) leads to

\begin{equation}\tilde{T}=Y',\label{160}\end{equation}

\begin{equation}\tilde{p}=\frac{5}{3}\,Y,\label{170}\end{equation}

\noindent and

\begin{equation}\tilde{\kappa}^{-1}=\frac{5}{3}\,Y-\frac{2}{3}      \frac{Y'^2}{Y''}.\label{180}\end{equation}

\noindent Clearly, if we know $Y(z)$, then we may construct a unique curve $(\tilde{p},\tilde{\kappa})$ (parameterized by $z$) from Eqs. (\ref{170}) and (\ref{180}).  The MIT group \cite{Ku2012} determined this curve not from any $Y(z)$, but from repeated density profile measurements, varying the trapping potential, the total number of atoms, and the temperature.  The precisely measured $(\tilde{p},\tilde{\kappa})$ curve, including error bars, is shown in the MIT group's Figure 1 \cite{Ku2012}, and directly displays a phase transition from a normal phase to a superfluid phase.  I refer to this curve as MIT1.  

\par
Given a function $Y(z)$, Eqs. (\ref{160}) and (\ref{180}) show that we may construct a single curve $(\tilde{T},\tilde{\kappa})$ (parameterized by $z$).  The MIT group \cite{Ku2012} found $\tilde{T}$ by integration over their $(\tilde{p},\tilde{\kappa})$ curve:

\begin{equation} \frac{T}{T_F}=\left(\frac{T}{T_F}\right)_i \mbox{exp}\left[\frac{2}{5}\int_{\tilde{p}_i}^{\tilde{p}} \frac{d\tilde{p}}{\tilde{p}-\frac{1}{\tilde\kappa}}\right],\label{190}\end{equation}

\noindent where the subscript ``$i$'' refers to a reference state chosen to lie in the virial regime.  The resulting $(\tilde{T},\tilde{\kappa})$ curve, including error bars, is shown in the MIT group's Figure 2a \cite{Ku2012}.  I refer to this curve as MIT2a.

\subsection{The Duke experiment}

\par
I also tested my fits against the DUKE1 data set \cite{Luo2007,Luo2009}, a comparison that required trap integration of the theory curve.  Trap integration was discussed in detail in \cite{Ruppeiner2014a}.  Define the trap integrated energy, entropy, and particle number $(E_t,S_t,N_t)$, respectively, by 

\begin{equation} E_t=\int\left[e(\vec{r}) + \rho(\vec{r})U(\vec{r})\right]\,d^3r,\label{210}\end{equation}

\begin{equation} S_t=\int s(\vec{r})\,d^3r,\label{220}\end{equation}

\noindent and

\begin{equation} N_t=\int \rho(\vec{r})\,d^3r,\label{230}\end{equation}

\noindent where $e(\vec{r})=E/V$ denotes the energy per volume, $s(\vec{r})=k_B z \rho$ denotes the entropy per volume, and $U(\vec{r})$ is the trapping energy per particle.  The trap integration is carried out at constant $T$ and constant total chemical potential $\mu_0=\mu+U(\vec{r})$.\footnote{Cao {\it et al.} \cite{Cao2011} found that to make a successful temperature calibration with the Duke data, it was necessary to use the entropy data labeled ''$S_{1200}^{*}/k_B$,'' \cite{Luo2009} that was corrected for the finite interaction strength in the weakly interacting gas.  This data column was argued to be the best measure of $S_t$, and I used it in this paper.}

\par
The Duke group used an optical trap with a Gaussian potential \cite{Luo2009}

\begin{equation} U(\vec{r})=U_0\left[1-\mbox{exp}\left(-2\,\tilde{r}^2\right)\right], \label{2050}\end{equation}

\noindent where $U_0=10\,\mu$K$\,k_B$, $\tilde{r}^2 = (x_1/a_1)^2 + (x_2/a_2)^2 + (x_3/a_3)^2$, $(x_1,x_2,x_3)$ are the spatial coordinates, and $\{a_1,a_2,a_3\}=\{52.20, 45.44, 1153.2\}\,\mu$m.  They scaled their energy  values with the trap Fermi energy (for a harmonic trap)

\begin{equation} E_F(N_t)= \hbar (\omega_1\omega_2\omega_3)^{1/3}(3 N_t)^{1/3}.\label{2060}\end{equation}

\noindent The two transverse and the axial trap frequencies were $\{\omega_1, \omega_2, \omega_3\}$ = $2\pi\{665,764,30.1\}$Hz, respectively, with $\omega_i=\sqrt{4U_0/ma_i^2}$ $(i=1,2,3)$, and with $m=6.015$ amu the mass of a $^6$Li atom.

\subsection{The fitting workflow}

Six free parameters, $(\alpha,y_0,y_1)$ and $(\gamma,\tilde{y}_0,\tilde{y}_1)$, define $Y_S(z)$ and $Y_H(z)$, respectively.  A seventh free parameter is $z_c$.  However, the two joining conditions Eqs. (\ref{110}) and (\ref{120}) leave a total of five independent free parameters.  Generally, a key measure of the quality of the fit of a function $w=w(x)$ to a set of $n$ data pairs $(x_i,w_i)$ is

\begin{equation} \chi^2 = \frac{1}{n}\sum\limits_i \left[\frac{(y-w_i)}{\sigma_i}\right]^2,\label{240}\end{equation}

\noindent where $\sigma_i$ is the standard deviation of the data pair $(x_i,w_i)$.  $\sigma_i$ reflects the uncertainty in both $x_i$ and $w_i$.  $\chi^2\sim 1$ corresponds to a high quality fit.

\par
The approximate value of $z_c$ is easily discerned from the MIT data.  With the exception of five data points corresponding to the intermediate ``critical rounding'' zone in which $z_c$ must lie, it is obvious which side of $z_c$ any data point lies on.  I found that moving $z_c$ within this intermediate zone did not cause substantial variation in the overall quality of the fits, so I simply kept $z_c$ roughly centered in the intermediate zone for each fit.

\par
I used four fit parameters: $(\alpha, \gamma, \tilde{y}_0, \tilde{y}_1)$.  For each quadruple of fit values, and the centralized $z_c$, I determined $( y_0,y_1)$ algebraically with the joining conditions Eqs. (\ref{110}) and (\ref{120}), leading to the determination of the full function $Y(z)$.  The fitting made primary use of MIT1, which was based on directly measured mechanical quantities.  I picked a grid of $(\alpha,\gamma)$ values, and for each grid point I varied $(\tilde{y}_0,\tilde{y}_1)$ to minimize $\chi^2$ for MIT1.  A contour diagram then visually reveals information about the best fits.

\par
The best overall primary fits must be consistent with the other two data sets.  For each fit in the primary MIT1 grid, I calculated $\chi^2$ when that fit is applied to DUKE1 and MIT2a [with no further variation in $(\tilde{y}_0,\tilde{y}_1)$].  Using results for all three data sets in selecting the best overall fit considerably narrows the uncertainty in the fit parameters.

\section{Results}

\begin{figure}
\begin{minipage}[b]{0.5\linewidth}
\includegraphics[width=2.7in]{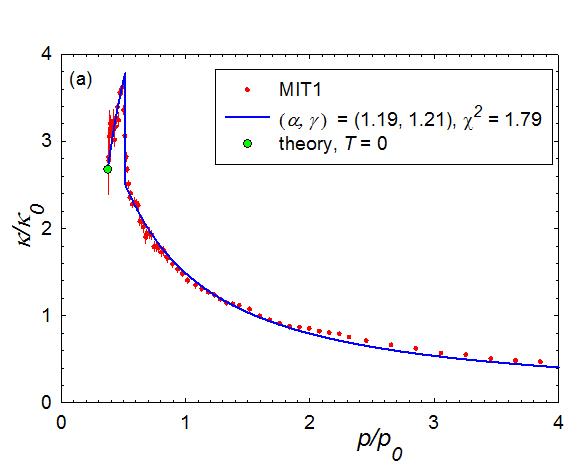}
\end{minipage}
\hspace{-0.2 cm}
\begin{minipage}[b]{0.5\linewidth}
\includegraphics[width=2.7in]{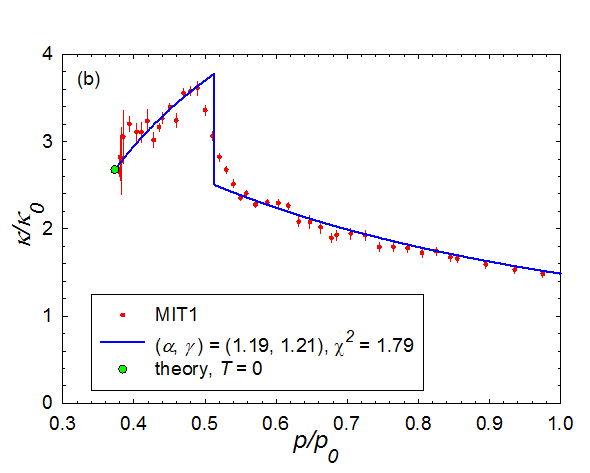}
\end{minipage}
\vspace{0.1cm}
\begin{minipage}[b]{0.5\linewidth}
\includegraphics[width=2.7in]{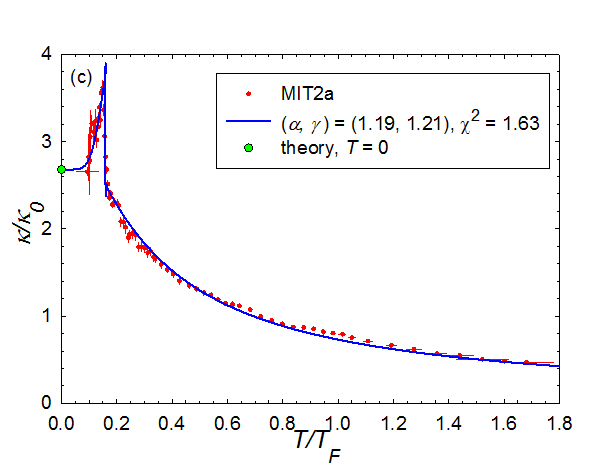}
\end{minipage}
\hspace{-0.2 cm}
\begin{minipage}[b]{0.5\linewidth}
\includegraphics[width=2.7in]{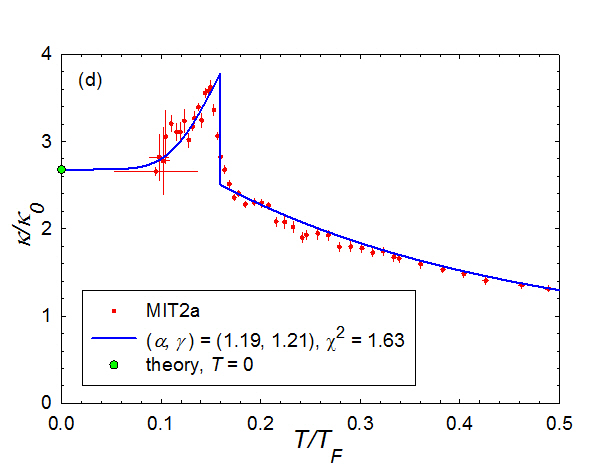}
\end{minipage}
\caption{The best overall fit, shown with MIT1 ($\chi^2=1.79$) and MIT2a ($\chi^2=1.63$).  The green dots correspond to $T=0$ on the theory curves, and yield the Bertsch parameter $\xi_B=0.373$.  This best overall fit has $z_c=0.652$.  The MIT group \cite{Ku2012} reported $\xi_B=0.376(4)$, and $z_c=0.73(14)$.}
\label{fig:2}
\end{figure}

In this section, I present the results, starting with the best overall fit.  This is followed by a discussion of how this fit was selected, and its uncertainties. I used the full spread of data points in MIT1 for the fitting, with some exceptions.  I omitted the five intermediate data points to the right of the peak point shown in Figure \ref{fig:2}(b).  I also omitted the five data points with the smallest values of $\tilde{p}$, since for these some trial theory curves had minimum $\tilde{p}$ larger than the experimental values.

\subsection{The best overall fit}

\par
Figure \ref{fig:2} shows the best overall fit, which has $\alpha=1.19$, $\gamma=1.21$, and $z_c=0.652$.  The green dots in Fig. \ref{fig:2} denote the zero-temperature points on the theory curves.  These points yield the value of the Bertsch parameter $\xi_B=0.373$, calculated from

\begin{equation}\xi_B=\tilde{p}=\tilde{\kappa}^{-1}\label{300}\end{equation}

\noindent at zero temperature.  $\xi_B$ gives the zero-temperature ratio of the energy per particle for the strongly interacting Fermi gas to that of the corresponding ideal Fermi gas \cite{Luo2009}.

\par
\begin{figure}[tpg]
\centering
\includegraphics[width=4.0in]{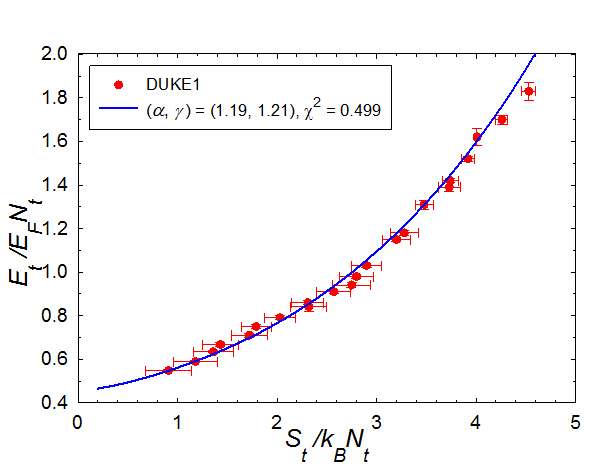}
\caption{The trap integrated best overall fit shown with DUKE1.  Trap integrating to compare with DUKE1 was an important element in selecting the best overall fit.  In this graph, $E_F$ denotes the Fermi energy of an ideal Fermi gas in a harmonic potential.}
\label{fig:3}
\end{figure}

Trap integrating this best overall fit curve to compare with DUKE1 yields Figure \ref{fig:3}.  As I discuss in the next subsection, DUKE1 offered major guidance for selecting the best overall fit.

\par
\begin{figure}
\begin{minipage}[b]{0.5\linewidth}
\includegraphics[width=2.7in]{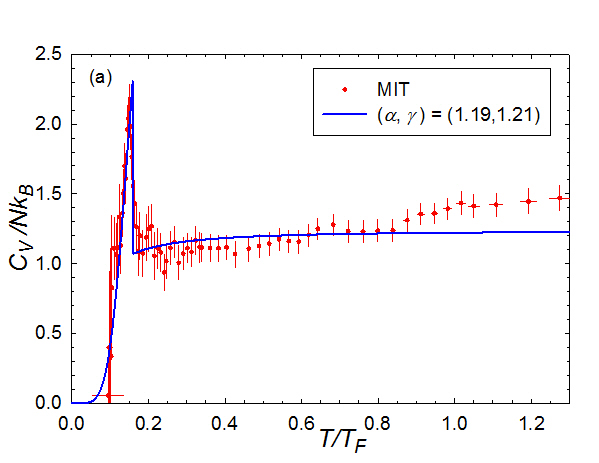}
\end{minipage}
\hspace{-0.2 cm}
\begin{minipage}[b]{0.5\linewidth}
\includegraphics[width=2.7in]{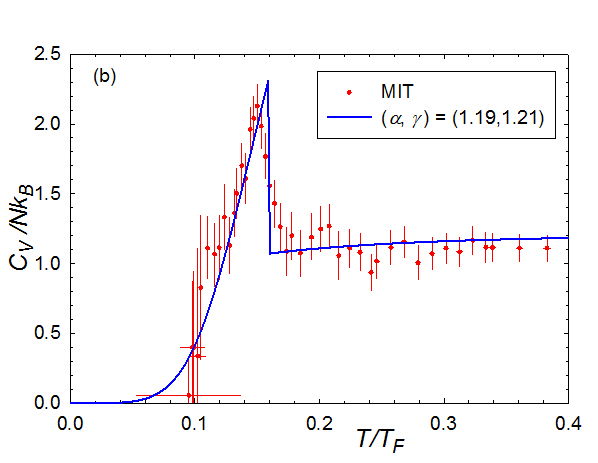}
\end{minipage}
\begin{minipage}[b]{0.5\linewidth}
\includegraphics[width=2.7in]{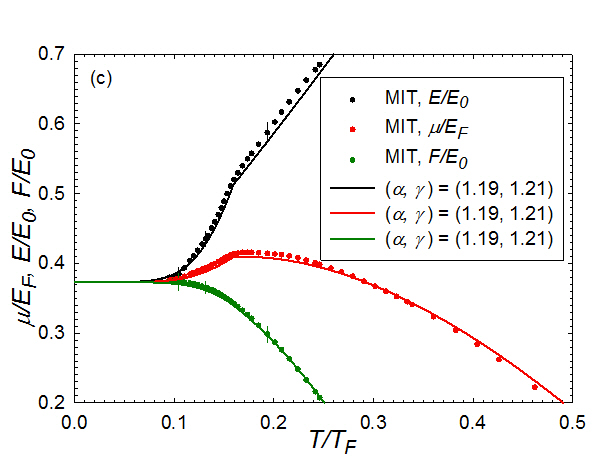}
\end{minipage}
\hspace{-0.2 cm}
\begin{minipage}[b]{0.5\linewidth}
\includegraphics[width=2.7in]{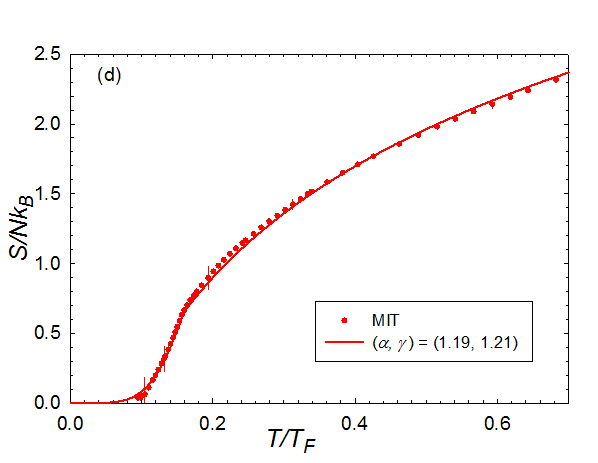}
\end{minipage}
\caption{Several functions of the temperature: (a) and (b) the heat capacity per particle, (c) the internal energy, the chemical potential, the Helmholtz free energy, and (d) the entropy per particle.}
\label{fig:4}
\end{figure}

Figure \ref{fig:4} compares curves derived from the best overall fit with the corresponding quantities of the MIT group \cite{Ku2012}.  These quantities are the heat capacity at constant volume per particle,

\begin{equation}\frac{C_V}{k_B N}=\frac{1}{k_B N}\left(\frac{\partial E}{\partial T}\right)_{V,N},\label{310}\end{equation}

\noindent the scaled internal energy $E/E_0$, where $E_0=3N\epsilon_F(\rho)/5$, the scaled chemical potential $\mu/E_F$, where $E_F=\epsilon_F(\rho)$, the scaled Helmholtz free energy $F/E_0$, where $F=E-TS$, and the scaled entropy $S/k_B N$ = $z$.

\par
While the curves in Fig. \ref{fig:4} show some nice agreement with experiment, systematic differences also show up.  Most significantly, the fit curve for the heat capacity in Fig. \ref{fig:4}(a) does not agree that well with experiment.  The experimental heat capacity data clearly reach a high temperature limit of $3/2$, characteristic of the monatomic ideal gas.  But the fit curve reaches a different limit because the value $\gamma=1.21$ corresponding to the best overall fit does not match the limiting ideal gas value $\gamma=1$.  Therefore, at this time the best overall fit does not match experiment in all respects, and deviations are most significant for the heat capacity.

\subsection{The fit analysis}

\begin{figure}
\begin{minipage}[b]{0.5\linewidth}
\includegraphics[width=2.7in]{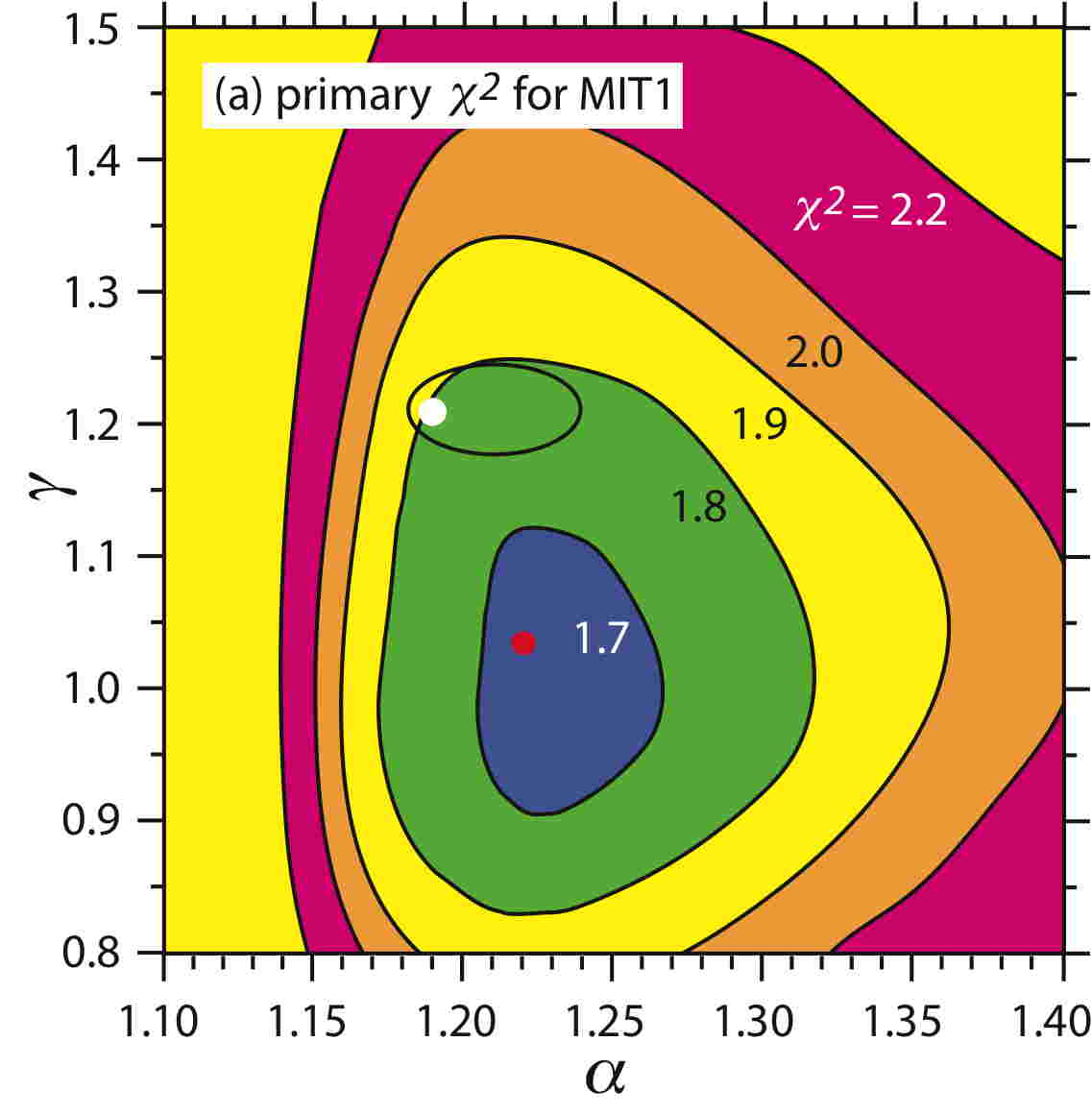}
\end{minipage}
\hspace{-0.2 cm}
\begin{minipage}[b]{0.5\linewidth}
\includegraphics[width=2.7in]{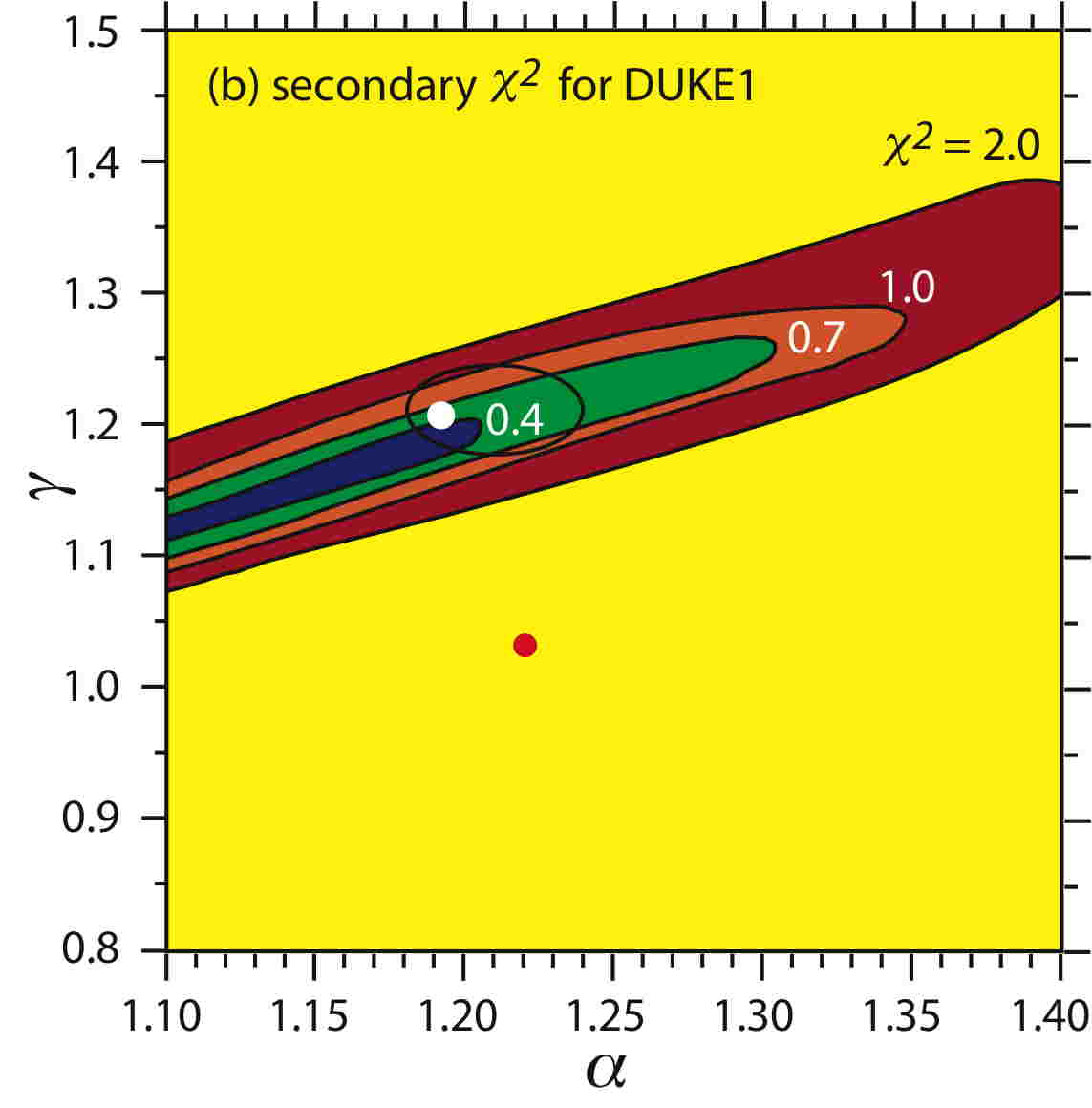}
\end{minipage}
\vspace{0.1cm}
\begin{minipage}[b]{0.5\linewidth}
\includegraphics[width=2.7in]{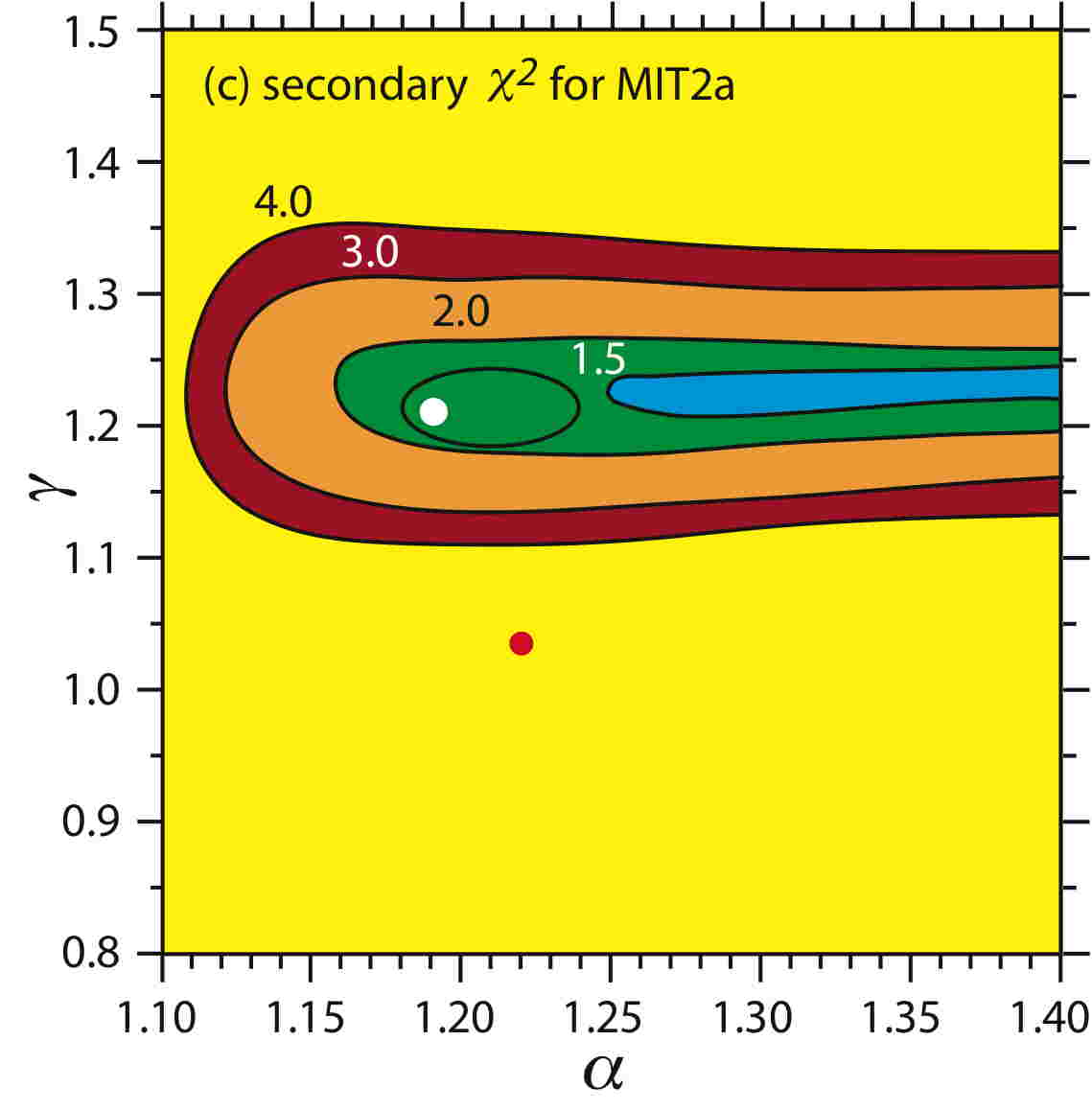}
\end{minipage}
\hspace{-0.2 cm}
\begin{minipage}[b]{0.5\linewidth}
\includegraphics[width=2.7in]{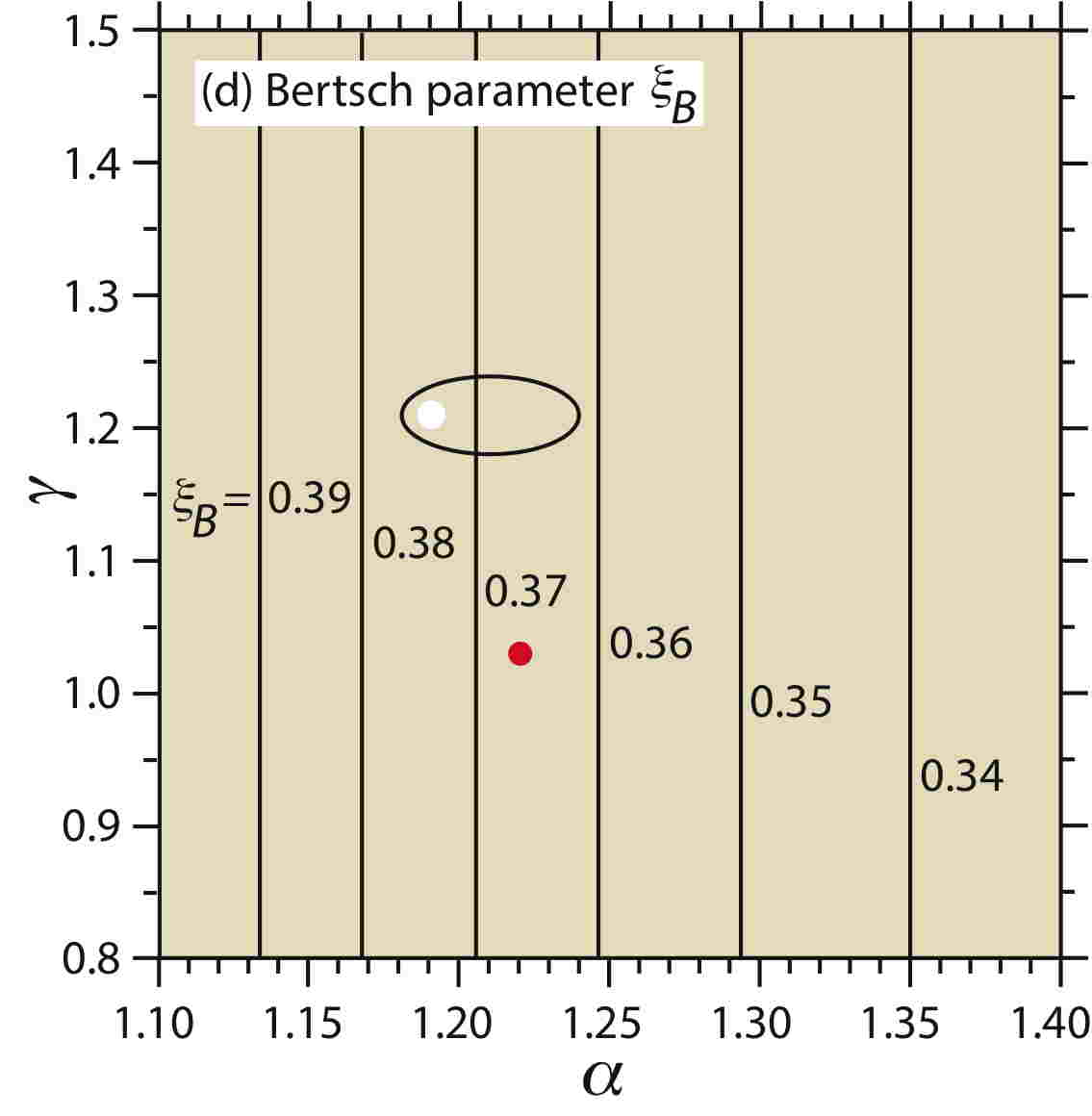}
\end{minipage}
\caption{Contour diagrams for the $(\alpha,\gamma,\chi^2)$ fit grid: (a) the primary MIT1 fit, (b) the secondary DUKE1 fit, (c) the secondary MIT2a fit, and (d) the Bertsch parameter grid.  The white dot in each diagram shows the best overall fit $(\alpha,\gamma)=(1.19,1.21)$, the red dot shows the best fit to MIT1 $(\alpha,\gamma)=(1.22,1.03)$, and the black oval shows the regime of the best overall fits.}
\label{fig:5}
\end{figure}

Figure \ref{fig:5}(a) shows the contour diagram for the primary $(\alpha,\gamma,\chi^2)$ MIT1 grid.  The best fit has $\{\alpha,\gamma,\chi^2\}=\{1.22,1.03,1.66\}$, and is indicated by a red dot.  Fig. \ref{fig:5}(a) shows that a broad range of values of $(\alpha,\gamma)$ produce reasonable fits to MIT1 ($\chi^2$ less than about $2$).
 
\par
To narrow the choice of best overall fit, I constructed secondary fit grids for DUKE1 and MIT2a, shown in Figs. \ref{fig:5}(b) and (c), respectively.  An acceptable fit has to match well all three data sets.  Inspection of the $\chi^2$ values in all the three fit grids revealed that nine fits performed well overall.  Table \ref{tab:1} shows six of these nine acceptable fits.  These nine fits are enclosed by the oval in Fig. \ref{fig:5}, indicating the rough uncertainty in $(\alpha,\gamma)$.  The best overall fit in the previous section is indicated by a white dot.  Statistics for the nine overall acceptable fits lead me to conclude that $\alpha=1.21(3)$, $\gamma=1.21(3)$, $z_c=0.69(2)$, $\xi_B=0.368(5)$, and $T_c/T_F=0.161(3)$.  The MIT group reported \cite{Ku2012} $z_c=0.73(13)$, $\xi_B=0.376(4)$, and $T_c/T_F=0.167(13)$, in good agreement with what was found here.

\par
In addition, there is a heat capacity jump $(c_v^--c_v^+)/c_v^+=1.14(4)$, where $c_v^-$ and $c_v^+$ denote the heat capacity at $z_c^-$ and $z_c^+$, respectively.  The BCS theory of superconductivity predicts a finite heat capacity jump of the type found here, with jump value $1.43$ \cite{Tinkham2004}.  The MIT group reports a lower bound jump value of $1.0_{-1}^{+4}$, which encompasses my value \cite{Ku2012}.

\begin{table}[h!]
\centering
\begin{tabular}{l|l|l|l|l|l|l}
\hline
			& A  			& B 			& C			& D 			& E 			& F 			 \\
\hline
$\alpha$		& $1.22$		& $1.20$		& $1.20$		& $1.21$		& $1.19$		& $1.21$		\\
$y_0$		& $0.21878$	& $0.22294$	& $0.22240$	& $0.22164$	& $0.22394$	& $0.22086$	\\
$y_1$		& $0.13162$	& $0.13359$	& $0.12912$	& $0.13400$	& $0.13372$	& $0.13261$	\\
$\gamma$	& $1.23$		& $1.21$		& $1.19$		& $1.21$		& $1.21$		& $1.20$		\\
$\tilde{y}_0$	& $0.12371$	& $0.12966$	& $0.12882$	& $0.12953$	& $0.13065$	& $0.12928$	\\
$\tilde{y}_1$	& $0.09690$	& $0.09572$	& $0.09642$	& $0.09436$	& $0.09557$	& $0.09450$	\\
$z_c$		& $0.689$		& $0.671$		& $0.684$		& $0.681$		& $0.652$		& $0.699$		\\
$\xi_B$		& $0.365$		& $0.372$		& $0.371$		& $0.369$		& $0.373$		& $0.368$		\\
$T_c/T_F$	& $0.161$		& $0.161$		& $0.156$		& $0.163$		& $0.159$		& $0.162$		\\
$\chi^2$ (a)	& $1.802$ 	&    $1.789$	& $1.759$ 	& $1.798$		& $1.791$		& $1.774$		\\
$\chi^2$ (b)	& $0.544$ 	&    $0.455$	& $0.436$ 	& $0.453$		& $0.499$		& $0.369$		\\
$\chi^2$ (c)	& $1.605$ 	&    $1.625$	& $1.788$	 	& $1.759$		& $1.629$		& $1.719$		\\

\hline
\end{tabular}
\caption{\label{tab:1} Parameters for six acceptable fits.  $\chi^2$ (a) corresponds to the primary MIT1 fits, $\chi^2$ (b) corresponds to the secondary DUKE1 fits, and $\chi^2$ (c) corresponds to the MIT2a fits.  Fit E was judged to be the best overall fit.}
\end{table}

\par
For each fit in the $(\alpha,\gamma)$ grid, I calculated a value for the Bertsch parameter $\xi_B$.  A contour diagram is shown in Fig. \ref{fig:5}(d).  Clearly, $\xi_B$ depends significantly on the value of $\alpha$, which is reasonable since $\alpha$ governs the low-$z$ behavior.  By contrast, effects of $\gamma$ on $\xi_B$ do not show up at all on the scale of the graph.

\par
I make one more observation.  If we multiply $T/T_F$ for MIT2a and $S_t$ for DUKE1 by the common factor 0.88, then the best fits for MIT2a and DUKE1 are brought close to that for MIT1, the red dot in Fig. \ref{fig:5}, which corresponds closely to a fitting function with the ideal gas limiting state ($\gamma=1$).  Logically, such a multiplicative factor might correspond to a temperature rescaling.  However, the temperature scale in this regime is known experimentally to within about 2\%, much less than the roughly 12\% deviation connected with my observation.  Nor is the improvement in statistics resulting from this observation, as measured by a reduction in $\chi^2$, significant.  So I do not claim any support for a temperature rescaling outside of this observation.

\section{Conclusions}
Unitary thermodynamics challenges experimentalists and theorists alike.  In this paper, I did a statistical analysis of two experimental efforts, one made at MIT, which collected $PVT$ data, and the other at Duke, which collected thermal data.   My fit function for the LDA was in two segments, reflecting fundamental physical differences above and below the phase transition at $z_c$.  I used a small set of independent fitting parameters: three parameters for the segment with entropy per particle $z$ above $z_c$, and one parameter for the segment with $z$ below $z_c$.

\par
A number of microscopic theories have been applied to the problem of unitary thermodynamics.  Examples are the viral series \cite{Liu2009, Nascimbene2010}, the T-matrix approach \cite{Haussmann2007}, and quantum Monte Carlo \cite{VanHoucke2012}.  However, a statistical analysis with the thermodynamic fit function here has advantages over comparisons with more fundamental microscopic theories.  First, microscopic theories rarely yield exact results, and their error can be difficult to assess.  Thus the $\chi^2$ minimization analysis used here is not usually available for microscopic theories, and it is such analysis that allows clear comparison between various data sets.  Second, no single microscopic theory can capture the full range of unitary thermodynamics, and it is necessary to mix and match theories to get the complete picture.

\par
My statistical analysis offers a clear comparison of data sets.  For example, the MIT group \cite{Ku2012} demonstrated that their $PVT$ data yields all of the thermodynamics, including the thermal properties.  The basis for their derivation was the fact that unitary thermodynamics follows a scaled equation of state.  Not as clear in their derivation, however, is the effect of error bars.  Although the error bars in the MIT data \cite{Ku2012} are mostly very small, particularly in the normal phase, they enlarge on calculating thermal properties.  Taken to the thermal regime, and trap integrated, these error bars do not appear to be significantly smaller than those found in the Duke experiment.   My analysis clearly brings this point out.

\par
My approach also offers straightforward trap integration.  Although there is considerable advantage to measuring the LDA directly, data averaged over a trap may be readily analyzed.  In the analysis of this paper, the MIT and the Duke experiments were complementary, and together revealed the complete thermodynamic picture.

\par
My analysis suggests some questions for future experiment and analysis: 1) For large entropy per particle $z$, does unitary thermodynamics approach the ideal gas thermodynamics?  Although the ideal gas limit will always be approached at the edge of the trap, where the particle density is small, and $z$ is large, it is not clear that this limit corresponds to unitary thermodynamics.  Experiments in larger traps, containing more atoms, would address this point.  2)  For small $z$, does unitary thermodynamics follow a power law characterized by an exponent $\alpha$.  More and better data below $T_c$ would lend insight into this question.  3) Could unitary thermodynamics be measured in spatial dimensions other than three, for example, in two dimensions?  Different dimensions correspond to different values of the exponents $a$ and $b$ in Eq. (3).  The approach in this paper could assist in addressing all of these questions.

\par
The main product of this paper was a good fit to the data sets of the MIT and Duke experimental groups, in the context of a thermodynamic geometric theory that is arguably correct.  All of the thermodynamic properties result. 

\section{Acknowledgements}

I thank G. Skestos for research support.  I also thank M. Zwierlein and M. Ku for sending their data tabulation, and for useful correspondence.  I thank J. Thomas and J. Joseph for useful correspondence, and for sending their computer code.  I also thank Masahito Ueda for helpful conversations.

\section{Appendix A: solution for $Y_H(z)$}

In this Appendix, I discuss the solution of the geometric equation for $Y_H(z)$ about the singular point $\mathcal{P}_0$ at $z\to\infty$, where we expect $R\to 0$, since this is the diffuse gas limit where interactions become weak.  Such singular points were discussed in detail in Appendix 1 of the first paper \cite{Ruppeiner2014a}, where it was argued that the appropriate form of the geometric equation and background subtraction is

\begin{equation} R = -\kappa\left[\frac{k_B T}{p}-\left(\frac{k_B T}{p}\right)_0\right]. \label{A10}\end{equation}

\noindent Here, the parentheses $()_0$ on the right-hand side of this equation mean evaluation at $\mathcal{P}_0$.

\par
The solution of Eq. (\ref{A10}) starts with a physically motivated series solution about $\mathcal{P}_0$.  In the first paper \cite{Ruppeiner2014a}, I used a critical phenomena style Puisseux series, and obtained a good fit to DUKE1.  However, it was not possible to fit MIT1 at high $z$ with such a solution.  MIT1 has tighter error bars and more high $z$ points than DUKE1, and thus poses a greater fitting challenge.  I tried another solution to Eq. (\ref{A10}) in the form of a perturbation about the ideal gas:

\begin{equation} Y_H(z)= \tilde{y}_0\,\mbox{exp}[2\gamma z/3]+ \tilde{y}_1\,\mbox{exp}[(2\gamma/3-1)z]+ \tilde{y}_2\,\mbox{exp}[(2\gamma/3-2)z]+\cdots, \label{A20}\end{equation}

\noindent where $\gamma$ and the first two series coefficients $\tilde{y}_0$ and $\tilde{y}_1$ may be picked freely.  The remaining series coefficients $\tilde{y}_i$ ($i\ge 2$) are determined in terms of $(\gamma,\tilde{y}_0,\tilde{y}_1)$ from the series solution of the differential equation for $Y_H(z)$, described below.  $\gamma=1$ corresponds to a limiting ($z\to\infty$) ideal gas solution (the Sackur-Tetrode equation, with an appropriate multiplier).

\par
From the methods of the first paper \cite{Ruppeiner2014a}, we may write:

\begin{equation} R = \frac{1}{\rho}\left[\frac{-10 Y(z) Y''(z)^2+5 Y(z) Y'(z)Y^{(3)}(z) + 5 Y'(z)^2 Y''(z)}{4 Y'(z)^3 - 10 Y(z) Y'(z) Y''(z)}\right]. \label{A30}\end{equation}

\noindent Define the series expansion parameter

\begin{equation}x=\mbox{e}^{-z}. \label{A40}\end{equation}

\noindent Eqs. (\ref{A20}) and (\ref{A30}) yield

\begin{equation} \begin{array}{lr} R=\displaystyle\left(\frac{15\tilde{y}_1}{8 \rho\gamma^2\tilde{y}_0}\right)x  \\ \hspace{1.0cm} 
 +\displaystyle\left(\frac{15[-15\tilde{y}_1^2+24\gamma\tilde{y}_1^2-16\gamma^2\tilde{y}_1^2+32\gamma^2\tilde{y}_0\tilde{y}_2]}{32\rho\gamma^4\tilde{y}_0^2}\right)x^2 +O(x^3). \end{array} \label{A50}\end{equation}

\noindent The definitions of $T$ and $p$ yield

\begin{equation}\displaystyle \frac{k_B T}{p}=\frac{\gamma}{\rho}-\left(\frac{3\tilde{y}_1}{2\rho\tilde{y}_0}\right)x -\left(\frac{3[-\tilde{y}_1^2+2\tilde{y}_0\tilde{y}_2]}{2\rho\tilde{y}_0^2}\right)x^2+ O(x^3). \label{A60}\end{equation}

\par
The series Eqs. (\ref{A50}) and (\ref{A60}) are related by the geometric equation Eq. (\ref{A10}).  Matching the zero'th order terms shows that the subtracter in Eq. (\ref{A10}) must be

\begin{equation}\left(\frac{k_B T}{p}\right)_0=\frac{\gamma}{\rho}.\label{A70}\end{equation}

\noindent The first-order terms in $x$ match, no matter what the values of the constants $\tilde{y}_0$ and $\tilde{y}_1$, provided that

\begin{equation}\kappa=\frac{5}{4\gamma^2}\label{A80}.\end{equation}

\noindent Matching second-order terms yields a linear algebraic equation for $\tilde{y}_2$, yielding its value uniquely in terms of $(\gamma,\tilde{y}_0,\tilde{y}_1)$.  Matching successively higher-order series terms now yields unique values for all of the remaining series coefficients $\tilde{y}_i\,(i\ge 3)$.

\par
Eq. (\ref{A10}) may be written as a third-order differential equation, which may be solved for any $z$, using the series for $Y_H(z)$ in Eq. (\ref{A20}) to generate initial conditions.  In practice, however, there was no need to solve the full differential equation because the series Eq. (\ref{A20}) converges very rapidly for all values $z>z_c\sim 0.7$.  Table \ref{tab:10} shows the series to increasing order for a solution corresponding closely to the best overall fit to MIT1.

\begin{table}[h!]
\centering
\begin{tabular}{l|c|c|c|l}
\hline
$n$ & $z=0.1$ & $z=0.3$ & $z=0.6$ & $z=1.0$ \\
\hline
0& $0.108329$ & $0.127125$ & $0.161607$ & $0.222554$ \\
1& $0.206349$ & $0.221301$ & $0.250299$ & $0.304427$ \\
2& $0.215279$ & $0.228327$ & $0.255201$ & $0.307460$ \\
3& $0.210579$ & $0.225299$ & $0.253636$ & $0.306811$ \\
4& $0.211694$ & $0.225887$ & $0.253861$ & $0.306874$ \\
5& $0.211670$ & $0.225877$ & $0.253858$ & $0.306873$ \\
6& $0.211588$ & $0.225848$ & $0.253852$ & $0.306872$ \\
\hline
\end{tabular}
\caption{\label{tab:10} Tabulation of results for the series for $Y_H(z)$, Eq. (\ref{A20}).  The series terminates with the $n$'th term, having coefficient $\tilde{y}_n$.  The parameter values $(\gamma,\tilde{y}_0,\tilde{y}_1)=(1.2,0.1,0.1)$.  These parameter values are close to the ones corresponding to the best fit.  Convergence to a value is rapid with increasing $n$, even for $z$ considerably less than $z_c\sim 0.7$.}
\end{table}

\section{Appendix B: virial expansion}

In this Appendix, the geometric equation is solved in the context of the viral expansion, which is generally employed for the regime of high $T$ and small $\rho$, a regime where approximately ideal gas behavior might be expected.  I show that a viral expansion is consistent with the geometric equation, and I calculate such an expansion for a particularly good limiting ideal gas ($\gamma=1$) fit to MIT1 from subsection 4.2.  Some comparison with virial expansions calculated by other means \cite{Ku2012, Liu2009, Ho2004b} is given, but a detailed discussion of the broader theoretical context is beyond the scope of this paper.

\par
The virial expansion for the free energy per volume $\phi$ is:\footnote{The series Eq. (\ref{B10}) is usually called a ``cluster expansion'' \cite{Pathria1996} rather than a ``virial expansion,'' but the later term is commonly employed in the literature of unitary thermodynamics.  I use the symbol $f$ for fugacity, since the usual symbol $z$ has a different meaning in this paper.}

\begin{equation}\phi\equiv\frac{p}{k_B T}=\frac{2}{\lambda^3}\left(b_1 f + b_2 f^2 + b_3 f^3 + O(f^4)\right),\label{B10}\end{equation}

\noindent where the expansion parameter

\begin{equation} f=\mbox{exp}\left(\beta\mu\right) \label{B20}\end{equation}

\noindent is the fugacity, $\beta=1/k_B T$, and

\begin{equation} \lambda=\frac{h}{\sqrt{2\pi m}}\,\beta^{1/2} \label{B30} \end{equation}

\noindent is the thermal wavelength.  Generally, the coefficients $b_i$, $i\ge 1$, may depend on $\beta$.  Here, they are taken to be constant.

\par
We have

\begin{equation} \phi = \frac{s}{k_B} - \beta e + \beta \mu \rho. \label{B35} \end{equation}

\noindent $\phi$ is naturally written as $\phi = \phi (\beta,h)$, where $h=-\beta\mu$, as in Eq. (\ref{B10}).  The energy per volume $e=-\phi_{,\beta}$, and the particle density $\rho=-\phi_{,h}$.  It is easy to show that to leading order in $f$, Eq. (\ref{B10}) yields standard ideal gas equations of state: $p=\rho k_B T$, $e=3\rho\,k_B T/2$, and $C_V/k_B N=3/2$.  Clearly, these equations of state are all independent of the value of $b_1$.

\par
In $(\beta,h)$ coordinates, the thermodynamic metric elements take the simple form $g_{\alpha\beta} = \phi_{,\alpha\beta}$ \cite{Ruppeiner1995}.  On using the series for $\phi$ in Eq. (\ref{B10}), and using Eq. (6) of reference \cite{Ruppeiner2014a} for $R$, we get

\begin{equation} R=-\frac{5\beta^{3/2}h^3}{16\sqrt{2}\,\pi^{3/2}m^{3/2}}\left[\frac{\,b_2}{\,b_1^2}+\frac{\,\left(-15\,b_2^2+8\,b_1 b_3\right)}{\,b_1^3}\,f+O\left(f^2\right)\right]. \label{B40} \end{equation}

\noindent Also,

\begin{equation} \displaystyle \frac{1}{\phi} = 
\displaystyle \frac{\beta^{3/2}h^3}{4\sqrt{2}\,\pi^{3/2}m^{3/2}}\left[\frac{1}{b_1 f}-\frac{\,b_2}{\,b_1^2}+\frac{\,\left(b_2^2-b_1 b_3\right)}{\,b_1^3}\,f+O\left(f^2\right)\right].  \label{B50}\end{equation}

\noindent Subtracting the series for $(1/\phi)_0$ computed from Eq. (\ref{A70}) removes the $1/b_1 f$ term in Eq. (\ref{B50}) and yields the singular part

\begin{equation} \displaystyle \left[\frac{1}{\phi}-\left(\frac{1}{\phi}\right)_0\right]=
\displaystyle \frac{\beta^{3/2}h^3}{4\sqrt{2}\,\pi^{3/2}m^{3/2}}\left[\frac{\,b_2}{\,b_1^2}+\frac{\,\left(-3b_2^2+2b_1 b_3\right)}{\,b_1^3}\,f+O\left(f^2\right)\right]. \label{B60}\end{equation}

\noindent Matching the zero'th-order terms in the series Eqs. (\ref{B40}) and (\ref{B60}) (with the second series multiplied by $-\kappa$, by Eq. (\ref{A10})) shows that $\kappa=5/4$, consistent with Eq. (\ref{A80}) for $\gamma=1$, as it must be, and that $b_1$ and $b_2$ may be set freely.  Matching the first-order series terms yields $b_3=2\,b_2^2/b_1$.  Matching higher-order series terms yields $b_4=505\,b_2^3/96\,b_1^2$, as well as all higher-order virial coefficients.

\par
Matching the leading term for $Y(z)$ in Eq. (\ref{A20}) to the leading term for the same quantity calculated with the virial expansion Eq. (\ref{B10}) yields

\begin{equation}b_1=\sqrt{\frac{6}{\pi e^5}} \,\tilde{y}_0^{-3/2},\label{B80}\end{equation}

\noindent where $e=2.71828$ is here, and in the next equation, the base of the natural logarithms.  Matching the leading terms in the two $R$ series Eqs. (\ref{A50}) and (\ref{B40}) leads to  

\begin{equation} b_2=-\frac{3}{2}\frac{\,b_1\tilde{y}_1}{\,e^{5/2}\tilde{y}_0}.\label{B90}\end{equation}

\par
Before calculating numbers, I comment on the sign of $b_2$.  For other Fermi systems, $R$ was found to be uniformly positive; see \cite{Ruppeiner2014} for review.  The best overall fit in this paper likewise has $R$ uniformly positive; see Figure \ref{fig:7}.  By Eq. (\ref{B40}) for $R$, we thus expect negative $b_2$.  Calculation with statistical mechanics for unitary thermodynamics yields positive $b_2$ \cite{Ku2012, Liu2009, Ho2004b}, at odds with the finding here.  A reconciliation in sign could be achieved if there were a zero crossing for $R$ somewhere in the normal phase.  However, a general argument has been made \cite{Ruppeiner2015} that such a zero crossing indicates a fundamental change in the character of the interparticle interactions, and it is not clear that this is in play here.  Whether or not a zero crossing by $R$ could possibly be consistent with the experimental data analyzed here is a question beyond the scope of this paper.  Let me add, however, that all of my fits in this paper have $\tilde{y}_0$ and $\tilde{y}_1$ both positive, so Eq. (\ref{B90}) also points to negative $b_2$.  The noninteracting Fermi gas also has negative $b_2$ \cite{Pathria1996}.

\par
Consider now the fit from subsection 4.2 with

\begin{equation} \{\alpha, \tilde{y}_0, \tilde{y}_1, z_c, \gamma\} = \{1.22, 0.137866, 0.087832, 0.841487, 1\}. \label{B95} \end{equation}

\begin{figure}[tpg]
\centering
\includegraphics[width=4.0in]{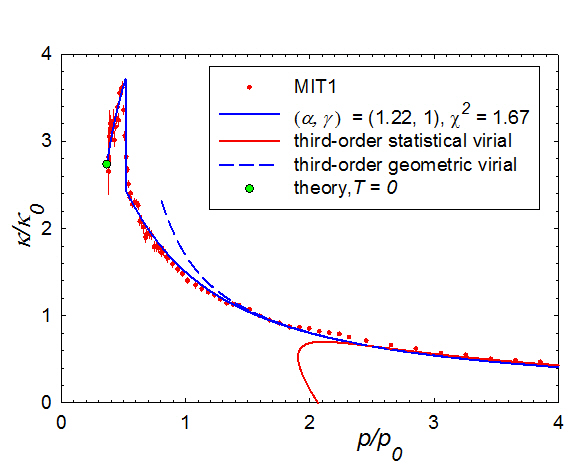}
\caption{A limiting ideal gas curve ($\gamma=1$) fit to all of MIT1, and the third-order geometric virial expansion resulting from this curve, with $\{b_1, b_2,b_3\}=\{2.21604, -0.173830, 0.027271\}$.  Also shown (red solid curve) is the curve from the third-order virial expansion from statistical mechanics, with $\{b_1, b_2, b_3\} = \{1, 3\sqrt{2}/8, -0.29095295\}$.}
\label{fig:6}
\end{figure}

\noindent  This fit to all of MIT1 ($\chi^2=1.67$) corresponds to a limiting ideal gas case ($\gamma=1$), and is near the best fit to MIT1.  Eqs. (\ref{B80}) and (\ref{B90}) yield $\{b_1, b_2\}=\{2.21604, -0.173830\}$, and the series solution to the geometric equation in $(\beta,h)$ coordinates yields $b_3=0.027271$.  Figure \ref{fig:6} shows MIT1, the curve from the fit in Eq. (\ref{B95}), and the curve generated by this fit on using just the first three terms in its virial series.

\par
The virial expansion coefficients have been calculated to third-order with statistical mechanics \cite{Ku2012, Ho2004b, Liu2009}: $\{b_1, b_2, b_3\} = \{1, 3\sqrt{2}/8, -0.29095295\}$.  The resulting curve is also shown in Fig. \ref{fig:6}, and it shows slightly better agreement with the very high temperature MIT1 data than does the curve from the thermodynamic geometric theory.  The later curve is stressed at high temperature trying to fit all of MIT1.  However, as the temperature is decreased, the thermodynamic geometric theory appears to yield a better converging third-order virial series than the one from statistical mechanics.

\par
Fitting curves with $\gamma=1$ to just to the normal phase MIT1 data yields good fits ($\chi^2\sim1.7$) for a broad range of $b_1$ values, including the statistical mechanical value $b_1=1$.  This insensitivity to $b_1$ results from the fact that the quantities plotted in Fig. \ref{fig:6} are independent of $b_1$ in the high temperature limit.  This fitting uncertainty for $b_1$ results in corresponding uncertainties in the values of the other virial coefficients, and makes a meaningful comparison with the statistical mechanical values difficult to do.  A detailed comparison is well beyond the scope of this paper.

\section{Appendix C: best fit function}

In this Appendix, I write an explicit expression for the best overall fit to the  energy $Y (z)$ per particle in units of the Fermi energy, presented in subsection 4.1.  I also write explicit expressions for a number of thermodynamic functions.

\par
$Y(z)$ is given in two analytic parts, $Y_S(z)$ and $Y_H(z)$, separated by a second-order phase transition at $z=z_c$, characterized by continuous $Y(z)$ and $Y'(z)$:
 
\begin{equation} Y(z) = \left\{ \begin{array}{llllll} Y_{S}(z) \mbox{  for } 0<z< z_c,\\   Y_{H}(z) \mbox{  for } z_c<z. \end{array}\right.\label{C10}\end{equation}

\noindent $Y_S(z)$ and $Y_H(z)$ may each be written as a series:

\begin{equation} Y_S(z)=y_0 + y_1 \, z^{\alpha} + y_2 \, z^{2 \alpha} + \cdots, \label{C20}\end{equation}

\noindent and

\begin{equation} Y_H(z)= \tilde{y}_0\,\mbox{exp}[2\gamma z/3] + \tilde{y}_1\,\mbox{exp}[(2\gamma/3-1)z] + \tilde{y}_2\,\mbox{exp}[(2\gamma/3-2)z] + \cdots. \label{C30}\end{equation}

\noindent The first seven coefficients for each series are shown in Table \ref{tab:20}, along with $(\alpha,\gamma,z_c)$.  Both these series have excellent convergence over the respective ranges of their functions.

\begin{table}
\begin{center}
\begin{tabular}{c|c|c}
\hline
$i$	& $y_i$								&	$\tilde{y}_i$					\\
\hline
\hline
$0$	& \hspace{10pt}	$0.223934789905$		& \hspace{10pt}	$0.130644611499$	\\
$1$	& \hspace{10pt}	$0.133726041923$		& \hspace{10pt}	$0.095571036084$	\\
$2$	& \hspace{10pt}	$0.010428751748$		& \hspace{10pt}	$0.007021911877$	\\
$3$	& \hspace{0pt}		$-0.000940774252$		& \hspace{0pt}		$-0.002938383443$	\\
$4$	& \hspace{10pt}	$0.000171419525$		& \hspace{10pt}	$0.000575606236$	\\
$5$	& \hspace{0pt}		$-0.000040543294$		& \hspace{0pt}		$-0.000019667129$	\\
$6$	& \hspace{10pt}	$0.000011033952$		& \hspace{0pt}		$-0.000024077262$	\\
\hline
\end{tabular}
\end{center}
\caption {The leading series coefficients for $Y_S(z)$ and $Y_H(z)$, which correspond to $\{\alpha,\gamma,z_c\}=\{1.19, 1.21,0.651793\}$, characterizing the best overall fit.}
\label{tab:20}
\end{table}

\par
Below, I list explicit expressions for a number of functions encountered in this paper:

\noindent The reduced pressure:

\begin{equation}  \tilde{p}=\frac{p}{p_0}=\frac{5}{3}\,Y(z). \end{equation}

\noindent The reduced temperature:

\begin{equation}  \tilde{T}=\frac{T}{T_F}=Y'(z). \end{equation}

\noindent The reduced chemical potential:

\begin{equation} \tilde{\mu}=\frac{\mu}{\epsilon_F}=  \frac{5}{3}\,Y(z) - z Y'(z).\end{equation}

\noindent The reduced inverse compressibility:

\begin{equation}  \tilde{\kappa}^{-1}=\frac{\kappa_0}{\kappa}=\frac{5}{3}\,Y(z) - \frac{2}{3}\frac{Y'(z)^2}{Y''(z)}. \end{equation}

\noindent The entropy per particle:

\begin{equation} \frac{S}{N k_B}=z. \end{equation}

\noindent The heat capacity:

\begin{equation} \frac{C_V}{N k_B}=\frac{Y'(z)}{Y''(z)}. \end{equation}

\noindent The reduced energy:

\begin{equation} \frac{E}{E_0}=\frac{5}{3}\,Y(z). \end{equation}

\noindent The reduced Helmholtz free energy:

\begin{equation}\frac{F}{E_0} = \frac{5}{3}\,\frac{F}{N \epsilon_{F}}=  \frac{5}{3} \left[Y(z)-z Y'(z)\right].  \end{equation}

\noindent The thermodynamic curvature:

\begin{equation} R=\frac{1}{\rho}\left[\frac{-10 Y(z) Y''(z)^2+5 Y(z) Y'(z) Y^{(3)}(z) +5 Y'(z)^2 Y''(z)}{4 Y'(z)^3-10 Y(z) Y'(z) Y''(z)}\right]. \end{equation}

\par
With these expressions in terms of $Y(z)$, it is possible to relate one property to any other.  Simply tabulate both properties in pairs (parametrized by $z$), interpolate one property versus the other, and plot.

\par
\begin{figure}[tpg]
\centering
\includegraphics[width=4.0in]{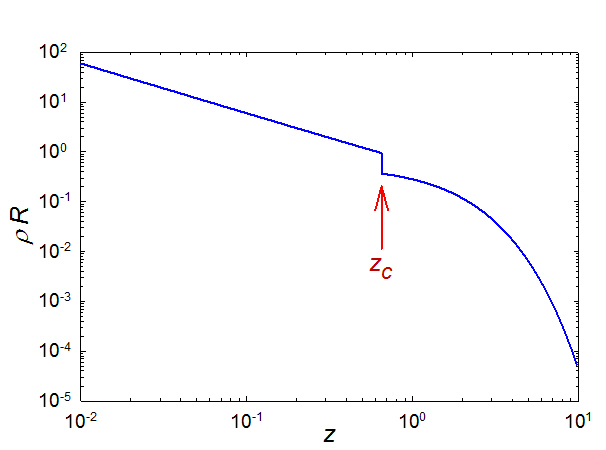}
\caption{$\rho R$ as a function of $z$.  $R$ is uniformly positive, and diverges to $+\infty$ as $T\to 0$.  $R$ shows a discontinuous jump at the phase transition $z_c=0.652$; $\rho R: 0.370 \to 0.948$.}
\label{fig:7}
\end{figure}

I conclude with a graph for the thermodynamic curvature $R$, shown in Figure \ref{fig:7}.  $R$ is seen to be uniformly positive for unitary thermodynamics over the full range of $z$, and diverges to $+\infty$ at $T\to 0$.  $R$ for the ideal Fermi gas is likewise uniformly positive, and diverges at $T\to 0$ \cite{Mrugala1990, Oshima1999}.

 \end{document}